\documentclass[a4paper,11pt]{article}
\usepackage{pos}

\newcommand{\vP}{\mathbf{P}}
\newcommand{\oto}[1]{\frac{1}{2\omega_{#1}}}
\newcommand{\ssum}{\raisebox{1pt}{\ensuremath{\scriptstyle\sum}}}
\newcommand{\sssum}{\raisebox{1pt}{\ensuremath{\scriptscriptstyle\sum}}}
\newcommand{\vx}{\mathbf{x}}
\newcommand{\vp}{\mathbf{p}}

\newcommand{\vk}{\mathbf{k}}
\newcommand{\gtt}{g_{2\to2}}
\newcommand{\gtf}{g_{2\leftrightarrow4}}
\newcommand{\gff}{g_{4\to4}}

\title{Towards four-pion effects in multi-hadron decays}

\author*{Rajnandini Mukherjee}
\author{Maxwell T. Hansen}

\affiliation[]{Higgs Centre of Theoretical Physics, School of Physics and Astronomy,\\
The University of Edinburgh, Edinburgh, UK}

\emailAdd{r.mukherjee@ed.ac.uk}
\emailAdd{maxwell.hansen@ed.ac.uk}

\abstract{The rigorous treatment of four-particle intermediate and final states poses a major challenge for lattice calculations of scattering and decay amplitudes, as well as long-distance matrix elements. As a step towards addressing these challenges, we present a new formalism that perturbatively relates two- and four-particle finite-volume energies and matrix elements to the couplings of the infinite-volume theory. Our method works at leading order in the two-to-two, four-to-four, and two-to-four couplings of the theory, while also capturing the leading finite-volume effects associated with two-to-two subprocess scattering in the four-particle sector. The result takes the form of a quantization condition which we implement numerically to produce a plot of volume-dependent energies for center-of-mass energies up to the six-particle threshold. The solutions exhibit a clear signature of two- and four-particle-like states and the avoided level crossings between them, which are particularly sensitive to the two-to-four coupling. We further discuss the implications of this formalism for quantifying the four-particle contributions in decay and transition amplitudes (e.g.~for hadronic $D$ decays).}

\FullConference{The 42nd International Symposium on Lattice Field Theory (LATTICE2025)\\
2-8 November 2025\\
Tata Institute of Fundamental Research, Mumbai, India\\}

\begin{document}
\maketitle

\section{Introduction}
Lattice QCD calculations are necessarily performed in a finite spatial volume and with imaginary (Euclidean) time. As a result, the spectrum of the numerically evaluated theory is discrete (rather than continuous as in infinite volume) and the computed correlation functions have decaying (rather than oscillatory) time dependence. As a result, lattice-determined quantities related to multi-particle dynamics differ significantly from their infinite-volume counterparts.

One of the leading approaches to resolving this mismatch is to treat finite-volume energies and matrix elements as intermediate observables. Extracting these, by fitting the Euclidean time dependence of correlators, removes all effects of the imaginary time. It then remains to relate the finite-volume quantities to physical scattering and decay amplitudes. This is achieved using dedicated formalisms, pioneered by L\"uscher~\cite{Luscher:1985dn,Luscher:1986pf,Luscher:1991cf} and Lellouch and L\"uscher~\cite{Lellouch:2000pv}, that treat finite-volume effects as a useful tool rather than as unwanted artifact. This has subsequently been extended to general two-particle and nearly general three-particle systems. See refs.~\cite{Briceno:2017max,Hansen:2019nir,Rusetsky:2019gyk,Mai:2021lwb,Hansen:2025fbj,Sharpe:2026mtt} for recent reviews.

For energies above the two-pion elastic threshold, the interpretation of the spectrum is further complicated by the fact that finite-volume QCD eigenstates are classified only by internal symmetries, e.g.~charge and flavour, and by rotational symmetries of the periodic box. In particular, there is no notion of asymptotic multi-particle states in a finite volume, and labels related to individual scattering channels are simply not meaningful. For example, the analog to the weak decay amplitude for $D \to \pi \pi$ is a finite-volume matrix element that receives contributions from all energetically allowed states with the same quantum numbers (e.g.~$\pi\pi, K \overline K, 4\pi, 6\pi, \eta \eta, \ldots$). Schematically,~\cite{Hansen:2012tf,Briceno:2014uqa,Briceno:2015csa,Hansen:2021ofl,Briceno:2021xlc}
\begin{align}
\left\langle E_n,L|H_W|D\right\rangle =
C_{\pi \pi}(E_n,L) A_{D \to \pi \pi}
+ C_{K \overline K}(E_n,L) A_{D \to K \overline K}
+ C_{4\pi}(E_n,L) A_{D \to 4\pi}
+ \ldots \,,
\label{eq:mixing}
\end{align}
where the coefficients $C_{x}(E_n,L)$ encode the mixing. The infinite-volume quantities of interest, $A_{D \to x}$, can be related to physical decay amplitudes by solving integral equations that incorporate all on-shell intermediate states~\cite{Jackura:2020bsk,Hansen:2020otl,Jackura:2022gib,Dawid:2023jrj,Dawid:2023kxu}. Perturbatively, higher-multiplicity states are expected to be suppressed at low energies. However, in the presence of final-state resonances, this need not hold, and all kinematically allowed contributions must be included to correctly interpret the matrix elements.

By now the formalism is well-established for weak decays into coupled two- and three-particle channels~\cite{Lellouch:2000pv,Kim:2005gf,Christ:2005gi,Meyer:2011um,Briceno:2012yi,Hansen:2012tf,Feng:2014gba,Briceno:2014uqa,Briceno:2015csa,Muller:2020wjo,Hansen:2021ofl,Briceno:2021xlc}, but the treatment of four-particle states is still in its infancy. As a first step toward such a framework, in these proceedings we introduce a perturbative approach to study four-particle effects in scattering and decay amplitudes. The remaining sections are organised as follows: in section~\ref{sec:framework} we describe the coupled-channel expansion for a system with two- and four-particle sectors. In section~\ref{sec:numerical_results} we present numerical results for the finite-volume spectrum of such a system, and discuss the implications for interpreting finite-volume matrix elements. We conclude with a summary and outlook in section~\ref{sec:summary}.

\section{Perturbative framework\label{sec:framework}}
Inspired by the approach of ref.~\cite{Hansen:2015zta}, we consider a QFT of scalar fields with Lagrangian density
\begin{gather}
\begin{aligned}
\mathcal{L}\, = \,&\frac{1}{2}\partial_\mu\phi\partial_\mu\phi + \frac{1}{2}m^2\phi^2 + \frac{\delta Z}{2}\partial_\mu\phi\partial_\mu\phi + \frac{\delta Z_m}{2}m^2\phi^2 + \frac{\gtt}{4!}\phi^4 + \frac{\gtf}{6!}\phi^6 + \frac{\gff}{8!}\phi^8,
\end{aligned}
\end{gather}
where we include four-, six-, and eight-particle interaction terms. This theory splits into even and odd sectors under the $\mathbb{Z}_2$ symmetry of $\phi\to-\phi$, and in this work we focus on the even sector involving $2\to2$, $2\to4$, $4\to 2$, and $4\to4$ interactions.
The names of the couplings are chosen to reflect the processes they contribute to at leading order in our expansion. Although we are inspired by two- and four-pion scattering and decays, we refer to the scatterers as generic particles here, in particular because we do not include the effects of isospin.

The counterterms for mass and wave-function renormalisation are tuned to reproduce the physical, infinite-volume particle mass ($m$) and the infinite-volume propagator with unit residue at the mass pole. We do not include counterterms for the interaction terms and work instead in a framework with bare couplings.

\subsection*{Finite volume}
Restricting the spatial volume to a periodic box of side length $L$ quantizes the spatial momenta, allowing us to write the scalar field as a sum of modes (in the time-momentum representation)
\begin{align}
\phi(\vx,\tau) = \frac{1}{L^3}\sum_{\vp} e^{+i\vp\cdot\vx}\,\tilde{\phi}_\vp(\tau),\quad \vp=\frac{2\pi}{L}\mathbf{n},\,\mathrm{with}\ \mathbf{n}\in\mathbb{Z}^3,
\end{align}
and the non-interacting, time-ordered Euclidean propagator as
\begin{align}
\tilde{\Delta}_\vp(\tau) \equiv \left\langle
\tilde{\phi}_\vp(\tau)\tilde{\phi}_{-\vp}(0)
\right\rangle_{g=0} = \frac{L^3}{2\omega_\vp}e^{-\omega_\vp|\tau|},\quad \mathrm{with}\;\omega_\vp=\sqrt{\vp^2 + m^2} \,. \label{eq:free_prop}
\end{align}

\subsection*{Coupled-channel expansion}
Consider a two-point correlation function of an operator $\mathcal{O}$ that couples to both two- and four-particle states in the finite volume:
\begin{equation}
C_L(E,\vP) = \int_{-\infty}^\infty dt e^{-iEt-\varepsilon t}\Big\langle \mathcal{O}_\vP(it)\mathcal{O}_\vP^\dagger(0)\Big\rangle_L \,,
\end{equation}
where we have rotated to Minkowski time and Fourier transformed to energy-momentum space. We emphasize that the finite-volume energies extracted in the lattice calculation do not depend on the metric signature and can equally well be viewed as the decay constants of the Euclidean exponents in the time-momentum representation or as the poles of the Minkowski energy-dependent correlation function. Thus our task is to analytically relate the poles of $C_L(E,\vP)$ to the couplings of the theory~\cite{Luscher:1986pf,Kim:2005gf,Hansen:2014eka}.

We next define $\delta C_L(E,\vP)$ as the difference between the full correlation function and a certain class of diagrams that have exponentially suppressed finite-volume effects (e.g.~those with only $t$-channel two-particle loops or six-particle loops). We then perform an all-orders diagrammatic expansion of $\delta C_L(E,\vP)$ while only identifying contributions that will shift the pole positions at order $O(g_{2\to2}, g_{2\leftrightarrow4}, g_{4\to4})$ and capture the leading finite-volume effects associated with two-to-two subprocess scattering in the four-particle sector.

We will present more details of the expansion in a future publication, but the result can be expressed compactly as
\begin{align}
\delta C_L(E,\vP)
&=
\raisebox{-0.65cm}{\includegraphics[scale=1.1]{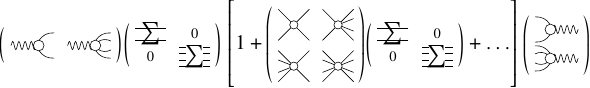}} \ ,\\
&= \sum_{n=0}^\infty AS[B S]^nA^\dagger = A \left[\frac{1}{S^{-1}-B}\right]A^\dagger.\label{eq:cc_exp}
\end{align}
The steps of identifying the two- and four-particle poles and factorizing the diagrams defining the correlator as shown  particularly follow the approach of refs.~\cite{Luscher:1986pf,Kim:2005gf,Hansen:2014eka,Briceno:2017tce,Blanton:2020gha}.

Here $S$ is a diagonal matrix of poles arising from $s$-channel two- and four-particle loops
\begin{gather}
S = \raisebox{-0.4cm}{\includegraphics[scale=1.0]{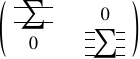}}
\equiv \begin{pmatrix}S_2 & 0 \\ 0 & S_4\end{pmatrix},
\end{gather}
with all incoming and outgoing momenta summed subject to the condition that the total momentum is $\vP$. Beginning with the two-particle component, in the time-momentum representation this is given by
\begin{align}
S_{2,\vP,L}(\tau) = \raisebox{-0.2cm}{\includegraphics[scale=1.1]{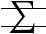}}
&\equiv \frac{1}{2!}\frac{1}{L^3}\sum_{\vk_1,\vk_2}\sum_{\vk_1',\vk_2'}\delta_{\vk_1+\vk_2,\vP} \ \raisebox{-0.7cm}{\includegraphics[scale=1.1]{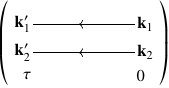}} \ , \\
&\overset{\text{\eqref{eq:free_prop}}}{=}\frac{1}{2!}\frac{1}{L^3}\sum_{\vk_1,\vk_2}\delta_{\vk_1+\vk_2,\vP}\oto{\vk_1}\oto{\vk_2}e^{-(\omega_{\vk_1}+\omega_{\vk_2})|\tau|}\,.
\end{align}
Analytically continuing and Fourier transforming to the energy-momentum representation, we reach
\begin{align}
S_2(E,\vP, L) &= \int_{0}^\infty dt \,e^{-iEt-\varepsilon t}S_{2,\vP,L}(it) \nonumber \\
&= \frac{1}{2!}\frac{1}{L^3}\sum_{\vk_i}\frac{\delta_{\sssum\vk,\vP}}{\prod_i2\omega_{\vk_i}}\frac{1}{\left(E - \ssum_i\omega_{\vk_i}
\right)} \,, \quad\text{with}\; i=1,2\,.
\end{align}
Here we do not include the integral from $-\infty$ to $0$, which contributes terms that are
smooth functions of $\vk_i$ and $E$, and thus have exponentially suppressed $L$-dependence
in our region of interest $2m<E<6m$.
These are absorbed into redefinitions of the couplings and endcap functions as we discuss below, and thus do not appear explicitly in our final result.
Note that the non-interacting two-particle energy levels correspond to poles at $E=\ssum_i \omega_{\vk_i} = \omega_{\vk_1}+\omega_{\vk_2}$ for any $\vk_1 + \vk_2 = \vP$.

Turning to $S_4$, this quantity contains diagrams involving four-particle loops, including those with all possible $2\to 2$ and $3 \to 3$ subprocesses. In the time-momentum representation, we have
\begin{gather}
\begin{aligned}
S_{4,\vP,L}(\tau) = \raisebox{-0.2cm}{\includegraphics[scale=1.0]{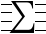}}
&\equiv \frac{1}{4!}\frac{1}{L^9}\sum_{\vk_1,\ldots,\vk_4}\sum_{\vk_1',\ldots,\vk_4'}\delta_{\ssum\vk,\vP} \\
&\qquad\times
\raisebox{-0.8cm}{\includegraphics[scale=1.0]{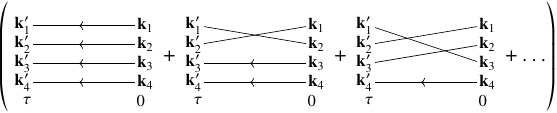}} \
.
\end{aligned}
\end{gather}

In this work we restrict attention the first two diagrams, which capture the leading finite-volume effects associated with two-to-two subprocess scattering in the four-particle sector.
Working in the time-momentum representation makes it straightforward to identify and evaluate the subleading term in the expansion: the subleading term generates a contribution of the form $\tau e^{- \sum_i \omega_{\vk_i} \tau}$, where the leading $\tau$ factor arises from expanding the exponentials of interacting energies to first order in the couplings. (Again, see ref.~\cite{Hansen:2015zta}.) Re-exponentiating this contribution and then Fourier transforming to energy-momentum space, we find
\begin{align}
S_4^{\mathrm{LO}}(E,\vP, L) = \frac{1}{4!}\frac{1}{L^9}\sum_{\vk_i}\frac{\delta_{\sssum\vk,\vP}}{\prod_i2\omega_{\vk_1}}\frac{c_{\vk}}{\left(E - \ssum_i\omega_{\vk_i} - \tilde{c}_{\vk} \frac{g_{2\to 2}}{L^3}
\right)}\,,
\quad \mathrm{with}\;i=1,\ldots,4 \,. \label{eq:S4LO}
\end{align}
The coefficients $c_{\vk}$ and $\tilde{c}_{\vk}$ encode symmetry factors specific to each momentum configuration. We have computed and incorporated these shifts for the lowest lying 50 poles in our numerical results below.

The poles of $S^{\text{LO}}_4$ give the leading-order four-particle spectrum in the absence of $2\leftrightarrow 4$ interactions given by
\begin{align}
E_n(L) = \sum_{i=1}^4\omega_{\vk_i} + \tilde{c}_{\vk}\frac{g_{2\to 2}}{L^3} + \cdots \,,
\end{align}
where the ellipsis denotes higher-order corrections in the couplings. For the threshold energy this reproduces the result of Huang and Yang~\cite{Huang:1957im}:
\begin{align}
E_0(L) = 4 m +
6 \frac{ g_{2 \to 2}}{8 m^2 L^3} + \cdots \,,
\end{align}
where $g_{2\to2}/(8 m^2 L^3)$ is the leading order energy shift of the two-particle energy and the factor of 6 counts the number of pairs of particles that can scatter.

We now turn to the two $L$-independent contributions to eq.~\eqref{eq:cc_exp}. First, $A^\dagger$ and $A$ are endcap functions -- given by vectors in the channel space -- that describe how the interpolating operator $\mathcal{O}$ creates and annihilates two- and four-particle states:
\begin{align}
A = \raisebox{-0.3cm}{\includegraphics[scale=1.0]{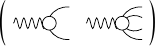}}
= \begin{pmatrix} A_{\mathcal{O} \to 2 \pi} & A_{\mathcal{O} \to 4 \pi} \end{pmatrix}.
\end{align}
These can be understood as the leading order short-distance contributions to matrix elements of the form $\langle 0|\mathcal{O}|2\pi\rangle$ and $\langle 0|\mathcal{O}|4\pi\rangle$.

Finally, $B$ is a matrix in channel space and contains all possible interactions vertices: $2\to2$, $2\to4$, $4\to 2$ and $4\to4$.
In full, all-orders finite-volume formalisms the quantity analogous to $B$ is initially defined with infinite-volume Bethe-Salpeter kernels. These are then converted to scheme-dependent $K$-matrices through a series of manipulations that involve absorbing smooth contributions arising from multi-particle loops.
In this work we have formally absorbed such smooth contributions because our definition of $S$ contains only simple two- and four-particle poles. But consistent with our aim to only include the leading order effects of the couplings on the finite-volume energies, we replace $B$ with its leading order expression, which is a matrix of the bare couplings:
\begin{align}
B = \raisebox{-0.65cm}{\includegraphics[scale=1.0]{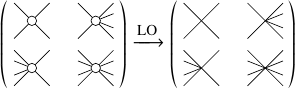}}
=\begin{pmatrix}g_{2\to 2} & g_{2\leftrightarrow 4} \\ g_{2\leftrightarrow 4} & g_{4\to4}\end{pmatrix}.
\end{align}

\subsection*{Regulating the two- and four-particle loops}
$S_2(E, \vP)$ and $S^{\text{LO}}_4(E, \vP)$ each involve a summation over an infinite tower of discrete momentum configurations. For practical numerical implementation, we regulate the sums by introducing a momentum cutoff (parameterised by a physical scale $\Lambda$) as well as a smooth damping factor (using a second dimensionful parameter $\alpha$) \cite{Luscher:1986pf,Kim:2005gf}. The regularised quantities are given by
\begin{align}
\mathcal{S}_2^{(\Lambda, \alpha)}(E,\mathbf{0},L) &= \frac{1}{2!}\frac{1}{L^3}\sum_{\vk_i}^\Lambda\frac{\delta_{\sssum\vk,\mathbf{0}}}{\prod_i2\omega_{\vk_i}}\frac{e^{\alpha(E-\ssum_i\omega_{\vk_i})}}{\left(E - \ssum_i\omega_{\vk_i}\right)},\quad i=1,2\,, \\
\mathcal{S}_4^{(\Lambda, \alpha)}(E,\mathbf{0},L) &= \frac{1}{4!}\frac{1}{L^9}\sum_{\vk_i}^\Lambda\frac{\delta_{\sssum\vk,\mathbf{0}}}{\prod_i2\omega_{\vk_i}}\frac{c_{\vk} e^{\alpha(E-\ssum_i\omega_{\vk_i})}}{\left(E - \ssum_i\omega_{\vk_i} - \tilde{c}_{\vk}\frac{g_{2\to2}}{L^3}\right)},\quad i=1,\ldots,4 \,,
\end{align}
where we will restrict attention to $\vP = \mathbf{0}$ for the remainder of this work.

The summation is now over the finite set of momenta $\{\vk\}$ such that each 3-momentum is constrained to $\vk_i \in \left[-\Lambda, \Lambda\right]^3$.
To make physical predictions, the bare couplings must be tuned as a function of the regulator parameters $\Lambda$ and $\alpha$ such that the predictions are invariant under changes in the regulator. Including the smooth damping factor seems to reduce the sensitivity of the bare parameters to the regulator in this tuning procedure.

\subsection*{Quantization condition and four-particle effects}

Returning to eq.~\eqref{eq:cc_exp} and dropping the $\mathbf{0}$ argument for brevity, the finite-volume spectrum for a given $L$ is given by all energies $E_n(L)$ at which the correlation function has a pole, which implies the quantization condition
\begin{align}
\operatorname{det}\left[\mathcal{S}^{(\Lambda,\alpha)}(E_n(L),L)^{-1} - B^{(\Lambda,\alpha)}\right] = 0\,. \label{eq:QC}
\end{align}
Here we have also given the coupling matrix $B$ a dependence on the regulator parameters, as discussed at the end of the previous section.
We note also that, because the couplings within $B$ have no momentum dependence, this quantization condition will only lead to interacting solutions for states that transform according to the trivial irrep of the relevant symmetry group, in particular as the $A_1^+$ irrep of the octahedral group with parity for the case of $\vP=\mathbf{0}$.

Note also that, for any energy $E$, and volume $L$, one can decompose the quantization condition matrix into its eigenvalues and eigenvectors as
\begin{align}
\mathcal{S}^{(\Lambda,\alpha)}(E,L)^{-1} -B^{(\Lambda,\alpha)} \equiv \begin{pmatrix} \mathbf{v}_1(E,L) & \mathbf{v}_2(E,L)\end{pmatrix}\begin{pmatrix}\lambda_1(E,L) & 0 \\ 0 & \lambda_2(E,L) \end{pmatrix}\begin{pmatrix}\mathbf{v}_1^T(E,L) \\ \mathbf{v}_2^T(E,L)\end{pmatrix} \,.
\end{align}

In the vicinity of a given solution one of the eigenvalues must vanish, and for concreteness we call it $\lambda_1$ ($\lambda_1(E_n(L), L) = 0$). Then as described in eq.~\eqref{eq:mixing}, and in analogy to the results of refs.~\cite{Hansen:2012tf,Briceno:2014uqa,Briceno:2015csa,Briceno:2021xlc},
the corresponding eigenvector gives the relative weights of the two- and four-particle components in the finite-volume state:
\begin{align}
\mathbf{v}_{1}(E_n, L)\propto\begin{pmatrix}C_{2\pi}(E_n, L) \\ C_{4\pi}(E_n, L)\end{pmatrix}. \label{eq:fpe}
\end{align}
The full value of the coefficients also depends on the energy derivative of the eigenvalue, but here we are mainly interested in the relative size of the two- and four-particle contributions, which is given by the ratio of the components of the eigenvector.

\section{Numerical results\label{sec:numerical_results}}

We now come to a first numerical evaluation of the results presented in the previous section. A reminder of the big picture: we are interested in studying the interacting finite-volume spectrum of such a system where the two- and four-particle sectors mix (allowed by a non-zero $g_{2\leftrightarrow4}$ coupling) in the energy region $2m<E<6m$. We work with total momentum $\vP=\mathbf{0}$, and study only the $A_1^+$ irrep.

\vspace{5pt}
\noindent
\begin{minipage}{0.48\textwidth}
\centering
\includegraphics[width=\linewidth]{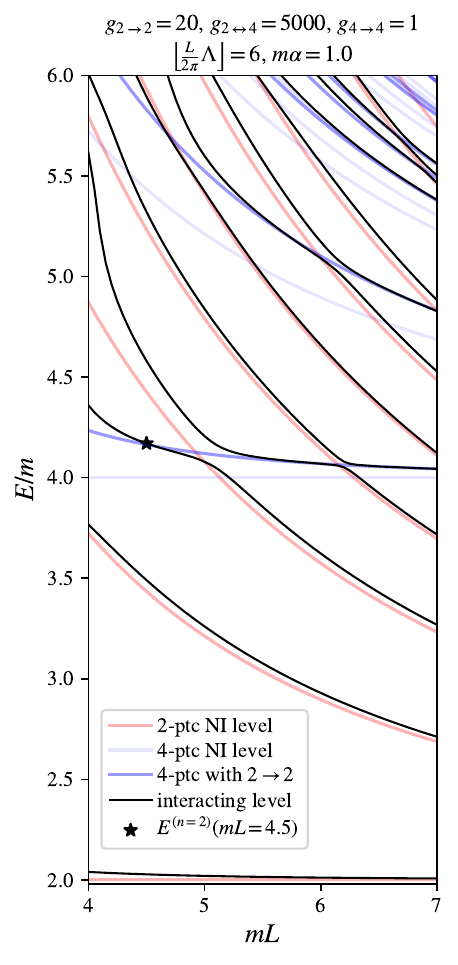}
\captionof{figure}{Interacting FV spectrum of a two- and four-particle system. Further explanation in the text.\label{fig:FV_spectrum}}
\end{minipage}
\hfill
\begin{minipage}{0.51\textwidth}
\setlength{\parindent}{1em}
Figure~\ref{fig:FV_spectrum} shows the interacting energy levels (black lines) obtained by numerically solving the quantization condition, eq.~\eqref{eq:QC}, for a fixed set of bare couplings and regulators. The light red curves are the non-interacting two-particle energy levels corresponding to the poles of $\mathcal{S}^{(\Lambda,\alpha)}_2$. The light (dark) blue levels correspond to the poles of $\mathcal{S}_4^{(\Lambda,\alpha)}$ in the absence (presence) of two-to-two interactions; i.e.~the difference between a light and dark blue level is the $O(g_{2\to2}/L^3)$ shift discussed in section~\ref{sec:framework}.

The non-interacting two- and four-particle levels exhibit true crossings, however the mixing of the sectors (due to $g_{2\leftrightarrow4}\neq 0$) causes level repulsions and we see this in the form of avoided level crossings. Consider the second excited state close to the four-particle threshold: naively one might argue that it behaves as a four-particle-ground-state-like level for $mL \lesssim 5$, and as a two-particle-excited-state-like level for $mL \gtrsim 5$.
We can make this comparison quantitative by evaluating eq.~\eqref{eq:fpe} for a particular $L$ value.
For example, taking the point marked by $\star$ in the figure, we find:
\begin{gather}
\mathbf{v}_{1}(E_2, L) = \begin{pmatrix} \scriptstyle{\mathtt{-0.99720554}} \\ \scriptstyle{\mathtt{-0.07470683}} \end{pmatrix}, \\
\implies
\frac{C_{4\pi}}{\sqrt{C_{2\pi}^2+C_{4\pi}^2}}
\sim 7.5 \%.
\end{gather}
\end{minipage}
\vspace{5pt}

\noindent In this case the four-particle contribution is enhanced relative to the expected volume suppression factor, but is still subdominant. We plan to investigate this dependence more in a forthcoming publication.

While we work to leading order in a power counting scheme where all three couplings are included, the two-to-two interactions have the least volume suppression and thus make the dominant contributions to the energy-level shifts. The magnitude of the shift within a given level depends on the multiplicity of allowed kinematic configurations. The effects of the four-to-four coupling are the most volume-suppressed while the two-to-four coupling governs the extent of mixing between the two sectors.

It is important to note that without lattice data it is not possible to predict the relative sizes of the bare couplings. In order to make a meaningful prediction for the size of four-particle effects in a given multi-hadron process, one must start by computing the finite-volume spectrum from lattice correlation functions, fit the spectrum to obtain the bare couplings, and then use this perturbative framework to make leading order estimates for four-particle contributions.

\section{Summary and outlook\label{sec:summary}}
In this work we have described a perturbative framework for studying finite-volume scattering systems where the two- and four-particle sectors mix. We used this to compute the finite-volume spectrum arising from two-to-two, two-to-four and four-to-four interactions and capture the behavior of and shifts in the interacting energies to leading order in the three bare couplings. Moreover, this framework predicts the relative size of the four-particle contributions to energies at above the four-particle threshold up to the six-particle threshold.

Some practical near-term extensions to this work are generalisations of the numerical implementation to non-zero total momentum ($\vP$), to other irreps of the relevant symmetry group, and to a realistic treatment of pions that incorporates isospin.
It would also be instructive to do a proof-of-principles lattice study using correlation functions of operators that overlap with two- and four-particle final states. One would then extract and fit the finite-volume spectrum to see how well lattice data can constrain the bare couplings of the theory, and whether meaningful predictions for four-particle effects can be made in such a setup.

\section*{Acknowledgments}
MTH and RM are supported by UKRI Future Leaders Fellowship MR/T019956/1. MTH is supported in part by UK STFC grants ST/X000494/1 and ST/T000600/1.

\bibliographystyle{JHEPmod}
\bibliography{ref}

@article{Hansen:2025fbj,
    author = "Hansen, Maxwell T.",
    title = "{Scattering on Periodic Lattices}",
    doi = "10.1007/978-3-031-90352-6_2",
    journal = "Lect. Notes Phys.",
    volume = "1041",
    pages = "43--88",
    year = "2025"
}

@article{Dawid:2023kxu,
	title        = {{Evolution of Efimov States}},
	author       = {Dawid, Sebastian M. and Islam, Md Habib E. and Brice\~no, Ra\'ul A. and Jackura, Andrew W.},
	year         = 2023,
	month        = 9,
	eprint       = {2309.01732},
	archiveprefix = {arXiv},
	primaryclass = {nucl-th}
}

@article{Dawid:2023jrj,
	title        = {{Analytic continuation of the relativistic three-particle scattering amplitudes}},
	author       = {Dawid, Sebastian M. and Islam, Md Habib E. and Brice\~no, Ra\'ul A.},
	year         = 2023,
	journal      = {Phys. Rev. D},
	volume       = 108,
	number       = 3,
	pages        = {034016},
	doi          = {10.1103/PhysRevD.108.034016},
	eprint       = {2303.04394},
	archiveprefix = {arXiv},
	primaryclass = {nucl-th}
}

@article{Jackura:2022gib,
	title        = {{Three-body scattering and quantization conditions from S-matrix unitarity}},
	author       = {Jackura, Andrew W.},
	year         = 2023,
	journal      = {Phys. Rev. D},
	volume       = 108,
	number       = 3,
	pages        = {034505},
	doi          = {10.1103/PhysRevD.108.034505},
	eprint       = {2208.10587},
	archiveprefix = {arXiv},
	primaryclass = {hep-lat},
	reportnumber = {JLAB-THY-22-3664}
}

@article{Hansen:2021ofl,
	title        = {{Decay amplitudes to three hadrons from finite-volume matrix elements}},
	author       = {Hansen, Maxwell T. and Romero-L\'opez, Fernando and Sharpe, Stephen R.},
	year         = 2021,
	journal      = {JHEP},
	volume       = {04},
	pages        = 113,
	doi          = {10.1007/JHEP04(2021)113},
	eprint       = {2101.10246},
	archiveprefix = {arXiv},
	primaryclass = {hep-lat}
}

@article{Hansen:2019nir,
    author = "Hansen, Maxwell T. and Sharpe, Stephen R.",
    title = "{Lattice QCD and Three-particle Decays of Resonances}",
    eprint = "1901.00483",
    archivePrefix = "arXiv",
    primaryClass = "hep-lat",
    doi = "10.1146/annurev-nucl-101918-023723",
    journal = "Ann. Rev. Nucl. Part. Sci.",
    volume = "69",
    pages = "65--107",
    year = "2019"
}

@article{Muller:2020wjo,
	title        = {{On the three-particle analog of the Lellouch-L\"uscher formula}},
	author       = {M\"uller, Fabian and Rusetsky, Akaki},
	year         = 2021,
	journal      = {JHEP},
	volume       = {03},
	pages        = 152,
	doi          = {10.1007/JHEP03(2021)152},
	eprint       = {2012.13957},
	archiveprefix = {arXiv},
	primaryclass = {hep-lat}
}

@article{Jackura:2020bsk,
	title        = {{Solving relativistic three-body integral equations in the presence of bound states}},
	author       = {Jackura, Andrew W. and Brice\~no, Ra\'ul A. and Dawid, Sebastian M. and Islam, Md Habib E. and McCarty, Connor},
	year         = 2021,
	journal      = {Phys. Rev. D},
	volume       = 104,
	number       = 1,
	pages        = {014507},
	doi          = {10.1103/PhysRevD.104.014507},
	eprint       = {2010.09820},
	archiveprefix = {arXiv},
	primaryclass = {hep-lat},
	reportnumber = {JLAB-THY-20-3272}
}

@article{Hansen:2020otl,
	title        = {{Energy-Dependent $\pi^+ \pi^+ \pi^+$  Scattering Amplitude from QCD}},
	author       = {Hansen, Maxwell T. and Brice\~no, Raul A. and Edwards, Robert G. and Thomas, Christopher E. and Wilson, David J.},
	year         = 2021,
	journal      = {Phys. Rev. Lett.},
	volume       = 126,
	pages        = {012001},
	doi          = {10.1103/PhysRevLett.126.012001},
	collaboration = {Hadron Spectrum},
	eprint       = {2009.04931},
	archiveprefix = {arXiv},
	primaryclass = {hep-lat},
	reportnumber = {CERN-TH-2020-147, JLAB-THY-20-3242}
}

@article{Blanton:2020gha,
	title        = {{Alternative derivation of the relativistic three-particle quantization condition}},
	author       = {Blanton, Tyler D. and Sharpe, Stephen R.},
	year         = 2020,
	journal      = {Phys. Rev. D},
	volume       = 102,
	number       = 5,
	pages        = {054520},
	doi          = {10.1103/PhysRevD.102.054520},
	eprint       = {2007.16188},
	archiveprefix = {arXiv},
	primaryclass = {hep-lat}
}

@article{Briceno:2017max,
	title        = {{Scattering processes and resonances from lattice QCD}},
	author       = {Briceno, Raul A. and Dudek, Jozef J. and Young, Ross D.},
	year         = 2018,
	journal      = {Rev. Mod. Phys.},
	volume       = 90,
	number       = 2,
	pages        = {025001},
	doi          = {10.1103/RevModPhys.90.025001},
	eprint       = {1706.06223},
	archiveprefix = {arXiv},
	primaryclass = {hep-lat},
	reportnumber = {JLAB-THY-17-2495, ADP-17-28-T1034},
	slaccitation = {%%CITATION = ARXIV:1706.06223;%%}
}

@article{Briceno:2017tce,
	title        = {{Relating the finite-volume spectrum and the two-and-three-particle $S$ matrix for relativistic systems of identical scalar particles}},
	author       = {Brice\~no, Ra\'ul A. and Hansen, Maxwell T. and Sharpe, Stephen R.},
	year         = 2017,
	journal      = {Phys. Rev. D},
	volume       = 95,
	number       = 7,
	pages        = {074510},
	doi          = {10.1103/PhysRevD.95.074510},
	eprint       = {1701.07465},
	archiveprefix = {arXiv},
	primaryclass = {hep-lat},
	reportnumber = {JLAB-THY-17-2400},
	slaccitation = {%%CITATION = ARXIV:1701.07465;%%}
}

@article{Briceno:2015csa,
	title        = {{Multichannel 0 $\to$ 2 and 1 $\to$ 2 transition amplitudes for arbitrary spin particles in a finite volume}},
	author       = {Brice\~no, Ra\'ul A. and Hansen, Maxwell T.},
	year         = 2015,
	journal      = {Phys. Rev. D},
	volume       = 92,
	number       = 7,
	pages        = {074509},
	doi          = {10.1103/PhysRevD.92.074509},
	eprint       = {1502.04314},
	archiveprefix = {arXiv},
	primaryclass = {hep-lat},
	reportnumber = {JLAB-THY-15-2009}
}

@article{Hansen:2014eka,
	title        = {{Relativistic, model-independent, three-particle quantization condition}},
	author       = {Hansen, Maxwell T. and Sharpe, Stephen R.},
	year         = 2014,
	journal      = {Phys. Rev. D},
	volume       = 90,
	number       = 11,
	pages        = 116003,
	doi          = {10.1103/PhysRevD.90.116003},
	eprint       = {1408.5933},
	archiveprefix = {arXiv},
	primaryclass = {hep-lat}
}

@article{Hansen:2012tf,
	title        = {{Multiple-channel generalization of Lellouch-Luscher formula}},
	author       = {Hansen, Maxwell T. and Sharpe, Stephen R.},
	year         = 2012,
	journal      = {Phys. Rev. D},
	volume       = 86,
	pages        = {016007},
	doi          = {10.1103/PhysRevD.86.016007},
	eprint       = {1204.0826},
	archiveprefix = {arXiv},
	primaryclass = {hep-lat},
	slaccitation = {%%CITATION = ARXIV:1204.0826;%%}
}

@article{Briceno:2012yi,
	title        = {{Moving multichannel systems in a finite volume with application to proton-proton fusion}},
	author       = {Briceno, Raul A. and Davoudi, Zohreh},
	year         = 2013,
	journal      = {Phys. Rev. D},
	volume       = 88,
	number       = 9,
	pages        = {094507},
	doi          = {10.1103/PhysRevD.88.094507},
	eprint       = {1204.1110},
	archiveprefix = {arXiv},
	primaryclass = {hep-lat},
	reportnumber = {NT-UW-12-05, NT@UW-12-05},
	slaccitation = {%%CITATION = ARXIV:1204.1110;%%}
}

@article{Luscher:1986pf,
	title        = {{Volume Dependence of the Energy Spectrum in Massive Quantum Field Theories. 2. Scattering States}},
	author       = {Luscher, M.},
	year         = 1986,
	journal      = {Commun. Math. Phys.},
	volume       = 105,
	pages        = {153--188},
	doi          = {10.1007/BF01211097},
	reportnumber = {DESY-86-034}
}

@article{Luscher:1985dn,
	title        = {{Volume Dependence of the Energy Spectrum in Massive Quantum Field Theories. 1. Stable Particle States}},
	author       = {L{\"u}scher, M.},
	year         = 1986,
	journal      = {Commun. Math. Phys.},
	volume       = 104,
	pages        = 177,
	doi          = {10.1007/BF01211589},
	reportnumber = {DESY-85-144}
}

@inproceedings{Sharpe:2026mtt,
    author = "Sharpe, Stephen R.",
    title = "{Three-particle scattering amplitudes from lattice QCD}",
    booktitle = "{42th International Symposium on Lattice Field Theory}",
    eprint = "2601.04147",
    archivePrefix = "arXiv",
    primaryClass = "hep-lat",
    month = "1",
    year = "2026"
}

@article{Briceno:2021xlc,
    author = "Brice{\~n}o, Ra{\'u}l A. and Dudek, Jozef J. and Leskovec, Luka",
    title = "{Constraining $1+\mathcal{J}\to 2$ coupled-channel amplitudes in finite-volume}",
    eprint = "2105.02017",
    archivePrefix = "arXiv",
    primaryClass = "hep-lat",
    reportNumber = "JLAB-THY-21-3365",
    doi = "10.1103/PhysRevD.104.054509",
    journal = "Phys. Rev. D",
    volume = "104",
    number = "5",
    pages = "054509",
    year = "2021"
}

@article{Hansen:2015zta,
	title        = {{Perturbative results for two and three particle threshold energies in finite volume}},
	author       = {Hansen, Maxwell T. and Sharpe, Stephen R.},
	year         = 2016,
	journal      = {Phys. Rev. D},
	volume       = 93,
	pages        = {014506},
	doi          = {10.1103/PhysRevD.93.014506},
	eprint       = {1509.07929},
	archiveprefix = {arXiv},
	primaryclass = {hep-lat}
}

@article{Mai:2021lwb,
	title        = {{Multi-particle systems on the lattice and chiral extrapolations: a brief review}},
	author       = {Mai, Maxim and D\"oring, Michael and Rusetsky, Akaki},
	year         = 2021,
	journal      = {Eur. Phys. J. ST},
	volume       = 230,
	number       = 6,
	pages        = {1623--1643},
	doi          = {10.1140/epjs/s11734-021-00146-5},
	eprint       = {2103.00577},
	archiveprefix = {arXiv},
	primaryclass = {hep-lat},
	reportnumber = {JLAB-THY-21-3327}
}

@article{Meyer:2011um,
	title        = {{Lattice QCD and the Timelike Pion Form Factor}},
	author       = {Meyer, Harvey B.},
	year         = 2011,
	journal      = {Phys. Rev. Lett.},
	volume       = 107,
	pages        = {072002},
	doi          = {10.1103/PhysRevLett.107.072002},
	eprint       = {1105.1892},
	archiveprefix = {arXiv},
	primaryclass = {hep-lat}
}

@article{Luscher:1991cf,
	title        = {{Signatures of unstable particles in finite volume}},
	author       = {L{\"u}scher, Martin},
	year         = 1991,
	journal      = {Nucl. Phys.},
	volume       = {B364},
	pages        = {237--251},
	doi          = {10.1016/0550-3213(91)90584-K},
	reportnumber = {DESY-91-052},
	slaccitation = {%%CITATION = NUPHA,B364,237;%%}
}

@article{Christ:2005gi,
	title        = {{Finite volume corrections to the two-particle decay of states with non-zero momentum}},
	author       = {Christ, Norman H. and Kim, Changhoan and Yamazaki, Takeshi},
	year         = 2005,
	journal      = {Phys. Rev.},
	volume       = {D72},
	pages        = 114506,
	doi          = {10.1103/PhysRevD.72.114506},
	eprint       = {hep-lat/0507009},
	archiveprefix = {arXiv},
	primaryclass = {hep-lat},
	reportnumber = {RBRC-530, SHEP-0520, CU-TP-1131},
	slaccitation = {%%CITATION = HEP-LAT/0507009;%%}
}

@article{Feng:2014gba,
	title        = {{Timelike pion form factor in lattice QCD}},
	author       = {Feng, Xu and Aoki, Sinya and Hashimoto, Shoji and Kaneko, Takashi},
	year         = 2015,
	journal      = {Phys. Rev.},
	volume       = {D91},
	number       = 5,
	pages        = {054504},
	doi          = {10.1103/PhysRevD.91.054504},
	eprint       = {1412.6319},
	archiveprefix = {arXiv},
	primaryclass = {hep-lat},
	reportnumber = {CU-TP-1206, KEK-CP-316, YITP-14-96},
	slaccitation = {%%CITATION = ARXIV:1412.6319;%%}
}

@article{Briceno:2014uqa,
	title        = {{Multichannel 1 $\rightarrow$ 2 transition amplitudes in a finite volume}},
	author       = {Brice{\~n}o, Ra\'ul A. and Hansen, Maxwell T. and Walker-Loud, Andr\'e},
	year         = 2015,
	journal      = {Phys. Rev.},
	volume       = {D91},
	number       = 3,
	pages        = {034501},
	doi          = {10.1103/PhysRevD.91.034501},
	eprint       = {1406.5965},
	archiveprefix = {arXiv},
	primaryclass = {hep-lat},
	reportnumber = {NT@WM-14-04, JLAB-THY-14-1909},
	slaccitation = {%%CITATION = ARXIV:1406.5965;%%}
}

@article{Lellouch:2000pv,
	title        = {{Weak transition matrix elements from finite volume correlation functions}},
	author       = {Lellouch, Laurent and L{\"u}scher, Martin},
	year         = 2001,
	journal      = {Commun. Math. Phys.},
	volume       = 219,
	pages        = {31--44},
	doi          = {10.1007/s002200100410},
	eprint       = {hep-lat/0003023},
	archiveprefix = {arXiv},
	primaryclass = {hep-lat},
	reportnumber = {CERN-TH-2000-091, CPT-2000-PE-3984, LAPTH-788-00},
	slaccitation = {%%CITATION = HEP-LAT/0003023;%%}
}

@article{Huang:1957im,
	title        = {{Quantum-mechanical many-body problem with hard-sphere interaction}},
	author       = {Huang, Kerson and Yang, C.N.},
	year         = 1957,
	journal      = {Phys. Rev.},
	volume       = 105,
	pages        = {767--775},
	doi          = {10.1103/PhysRev.105.767}
}

@article{Rusetsky:2019gyk,
    author = "Rusetsky, Akaki",
    title = "{Three particles on the lattice}",
    eprint = "1911.01253",
    archivePrefix = "arXiv",
    primaryClass = "hep-lat",
    doi = "10.22323/1.363.0281",
    journal = "PoS",
    volume = "LATTICE2019",
    pages = "281",
    year = "2019"
}

@article{Kim:2005gf,
    author = "Kim, C. h. and Sachrajda, C. T. and Sharpe, Stephen R.",
    title = "{Finite-volume effects for two-hadron states in moving frames}",
    eprint = "hep-lat/0507006",
    archivePrefix = "arXiv",
    reportNumber = "UW-PT-05-16, SHEP-0518",
    doi = "10.1016/j.nuclphysb.2005.08.029",
    journal = "Nucl. Phys. B",
    volume = "727",
    pages = "218--243",
    year = "2005"
}

\end{document}